\documentclass[superscriptaddress,
twocolumn,showpacs,preprintnumbers,amsmath,amssymb]{revtex4}
\usepackage[dvips]{graphicx}
\usepackage{epsf}
\usepackage{epsfig}
\usepackage{latexsym}
\usepackage{amssymb}
\usepackage{amsfonts,amsbsy}
\usepackage{amsmath}
\usepackage{dcolumn}
\usepackage{bm}
\usepackage{pifont}
\begin{document}
\title{Trafficlike collective movement of ants on trails: absence of 
jammed phase}

\author{Alexander John}%
 \affiliation{%
Institut  f\"ur Theoretische  Physik, Universit\"at 
zu K\"oln D-50937 K\"oln, Germany.
}%

\author{Andreas Schadschneider}%
\affiliation{%
Institut  f\"ur Theoretische  Physik, Universit\"at 
zu K\"oln D-50937 K\"oln, Germany.
}%
\affiliation{%
Interdisziplin\"ares Zentrum f\"ur komplexe Systeme,
University of Bonn, Germany.}

\author{Debashish Chowdhury}
\affiliation{%
Department of Physics, Indian Institute of Technology,
Kanpur 208016, India.
}%
\affiliation{%
Max-Planck Institute for Physics of Complex Systems, 01187 Dresden,
Germany.
}%
\author{Katsuhiro Nishinari}
\affiliation{%
Department of Aeronautics and Astronautics,
School of Engineering, University of Tokyo,
Hongo, Bunkyo-ku, Tokyo 113-8656, Japan.
}%
\affiliation{%
PRESTO, Japan Science and Technology Agency, Tokyo 113-8656, Japan}

\date{\today}
\begin{abstract}
We report experimental results on unidirectional 
traffic-like collective movement of ants on trails. Our work is 
primarily motivated by fundamental questions on the collective 
spatio-temporal organization in systems of interacting motile constituents 
driven far from equilibrium. Making use of the analogies with vehicular 
traffic, we analyze our experimental data for the spatio-temporal 
organisation of the ants on the trail. From this analysis, we extract 
the flow-density relation as well as the distributions of velocities 
of the ants and distance-headways. Some of our observations are 
consistent with our earlier models of ant-traffic, which are appropriate 
extensions of the asymmetric simple exclusion process (ASEP). 
In sharp contrast to highway traffic and most other
transport processes, the average velocity of the 
ants is almost independent of their density on the trail. Consequently, 
no jammed phase is observed. 

\end{abstract}
\maketitle

Ants form large trail systems \cite{holl90} which share many features
of vehicular transportation networks. Emergence of the trail pattern
has received some attention in the literature
\cite{holl90,camazine,gan-topology}.  Single trails are often stable
for hours or days and can be considered the analogs of highways.
Threfore, the collective movement of ants on trails (from now onwards,
referred to as ``ant-traffic'') is analogous to vehicular traffic on
highway networks \cite{css,polrev}. The social behavior of ants also
indicates the possibility that biological evolution has optimized
ant-traffic. Surprisingly, despite its striking similarities with
vehicular traffic, the {\em collective} properties of ant-traffic have
not been studied experimentally until recent years.

The pioneering experiments on ant-traffic \cite{burd1} and all the
subsequent related works \cite{burd2,couzin,johnson,duss05,john-swarm}
used {\em bidirectional} trails where the nature of flow is dominated by the
head-on encounters of the ants coming from opposite directions
\cite{burd2,couzin,john-swarm}.  But, in vehicular traffic, where flows
in opposite directions are normally well separated and head-on
collisions can occur only accidentally, the spatio-temporal
organization of the vehicles in each direction is determined by the
interactions of the vehicles moving in the {\em same} direction. Therefore,
in order to investigate the similarities and differences between
vehicular traffic and ant-traffic, we have collected and analyzed data
on {\it unidirectional} traffic of ants on a natural trail using
methods adapted from traffic engineering \cite{partha,may,kerner}
and the theory of stochastic processes \cite{mahnke}.

All the experimental data reported here have been collected on a
natural trail of monomorphic ant species \textit{Leptogenys
  processionalis} \cite{gan-topology}. This choice ensured that
all the ants have the same body size and exhibit identical behavioral
responses. Moreover, we maintained the natural situation so that the
true features of ant-traffic could be captured by our video
recordings. Furthermore, we focussed on a particular section of the
trail which had neither crossings nor branching which would be the
analogs of ramps in vehicular traffic \cite{may,kerner}.  Thus, being
far from nest and the food as well as from intersections, this segment
mimics an effectively infinite linear trail \cite{schuetz,css}. The
shape of the observed section of the trail remained unaltered for
several hours and, therefore, we collected each data set continuously
for about 13 minutes. We have verified that during this time the flow
can be considered to be stationary and is not disturbed by external
factors.  Finally, we compared the data recorded at ten different
trails of the same type and found that our conclusions drawn from
these are generic (at least, for the traffic of the ant species used
in our studies) \cite{johnphd}.

\begin{figure}[h]
\begin{center}
\includegraphics[width=0.48\textwidth]{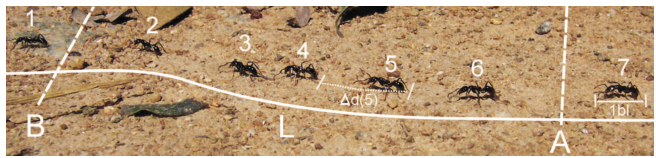}\
\end{center}
\vspace{-0.4cm}
\caption{A snapshot of the observed trail section. We obtained 
its length $L=17~$bl in the units of the body-length 
(bl) of a single ant. For the observed species one finds 1~bl\ $\approx 18~$mm.}
\label{fig-trail}
\end{figure}

One of the distinct behavioral characteristics of individual ants in 
the ant-traffic is the absence of overtaking. Although some ants 
(temporarily) left the trail and were passed by succeeding ones, we 
never observed any ant speeding up in order to overtake some other 
ant in front. We exploited this observation to develop a convenient 
method for our data analysis which we'll explain below. 

The actual length $L$ of the observed section of the trail between the 
two points marked A and B in Fig.~\ref{fig-trail} has been estimated 
by averaging over the paths of individual ants which passed through 
this section. Putting equispaced marks on a transparency mounted on 
the video screen, we obtained $L=17~$bl, in the units of the body-length 
(bl) of a single ant, where 1~bl\ $\approx 18~$mm for the ant species 
\textit{Leptogenys processionalis} used in our study. 

\begin{figure}[tb]
\begin{center}
\includegraphics[width=0.4\textwidth]{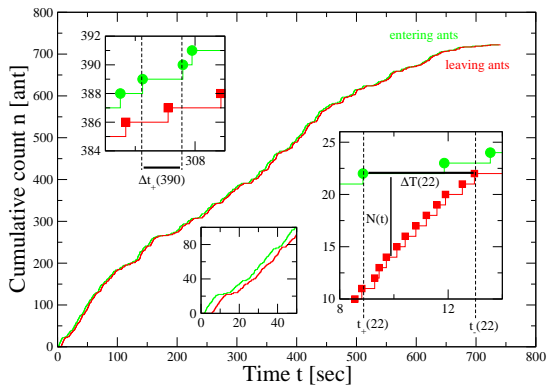}
\end{center}
\caption{Figure illustrating the technique employed for data extraction. 
  The cumulative count of the ants which have entered $n_{+}(t)$
  ($\bullet$) and left $n_{-}(t)$ $(\blacksquare)$ the trail section
  between A and B.  The right inset shows the travel time $\Delta T$
  for the 22th ant.  On the left inset the time-headway $\Delta t_{+}$
  of the $390th$ ant is shown.  }
\label{fig-ccount}
\end{figure}

Since no overtaking takes place, ants can be uniquely identified by
the ordered sequence in which they enter the observed section of the
trail, i.e.\ they follow a FIFO-principle (first-in-first-out).
Suppose, the $n$-th ant enters the section at $A$ at time $t_{+}(n)$ and
leaves the section at $B$ at time $t_{-}(n)$.  An efficient tool for
analyzing such data is the {\em cumulative plot}
(Fig.~\ref{fig-ccount}) (\cite{partha}); it shows the numbers $n_+(t)$
and $n_-(t)$ of ants which have passed the point $A$ and $B$,
respectively, up to time $t$. The two resulting curves, which are
sometimes called {\em arrival function} and {\em departure function}
can be obtained by inverting $t_{+}(n)$ and $t_{-}(n)$, respectively.

This allows us to extract basic data in a very efficient way 
\footnote{Data points on the arrival function without corresponding
point on the departure function, corresponding to ants leaving the trail
within the observed section, have been deleted. This happened in less
than 1\% of the cases within the observation interval.}. 
The travel time $\Delta T(n)$ of the $n$-th ant in the section 
between the points A and B is given by 
\begin{equation}
\Delta T(n) =  t_{-}(n)-t_{+}(n) 
\label{delt-def}
\end{equation} 
and the {\it time-averaged} speed of the $n$-th ant during the period 
$\Delta T(n)$ is given by  
\begin{equation}
v(n) =\frac{L}{\Delta T(n)}
\label{vel-def}
\end{equation}
The time-headway of two succeeding ants can be obtained easily
at the entrance and exit points $A$ and $B$ (Fig.~\ref{fig-ccount}, 
left inset). Since $v(n)$ is, by definition (Eqn.~(\ref{vel-def})), 
the time-averaged velocity $v(n)$ of the $n$-th ant, the distance-headway 
between the $n$-th ant and the ant in front of it is given by 
\begin{eqnarray}
\Delta d(n)=\Delta t_{+}(n)~ v(n-1) \,,  \nonumber\\
\Delta t_{+}(n)=t_{+}(n)-t_{+}(n-1)\,. 
\label{dh-def}
\end{eqnarray}
Entry and exit of each ant changes the instantaneous number $N(t)$ 
of the ants in the trail section between A and B by one unit
(Fig.~\ref{fig-ccount}, right inset). Therefore $N(t)$ fluctuates, 
but stays constant in between two events of entry or exit. Sorting 
the counts of these events by time one obtains a chronological list 
$\left\lbrace t_{i}\right\rbrace=\left\lbrace t_{\pm}(n)\right\rbrace$ 
of the changes of the instantaneous particle number  
\begin{equation}
N(t)=n_{+}(t)-n_{-}(t)=const. \quad  \text{while } t\in[t_{i},t_{i+1}[.
\end{equation}
In order to estimate the local density which is experienced by the $n$-th 
ant during the time interval $\Delta T(n)$ it spends within the observed 
trail section we first determine the average number of ants in the same 
section during the time interval $\Delta T(n)$:
\begin{eqnarray}
  \left\langle N\right\rangle_{t(n)} & =& \frac{1}{\Delta T(n)} 
\sum_{t_{i}=t_{+}(n)}^{t_{i}<t_{-}(n)}  N(t_{i})(t_{i+1}-t_{i}) 
\label{avN}
\end{eqnarray}
The (dimensionless) density $\rho(n)$ affecting the movement of the $n$-th 
ant is given by
\begin{equation}
\rho(n) =\dfrac{\left\langle N\right\rangle_{t(n)}}{N_{\text{max}}}
        =\dfrac{\tilde{\rho}(n)}{\tilde{\rho}_{\text{max}}}
\qquad\text{with}\quad
\tilde{\rho}(n)=\dfrac{\left\langle N\right\rangle_{t(n)}}{L}\, , 
\end{equation}
where $N_{\text{max}}=17 = L/(1~\text{bl})$ and $\tilde{\rho}_{\text{max}}=
N_{\text{max}}/L$. Our empirical data for $\rho$ are in the interval
$[0,0.8]$. The instantaneous particle numbers and the single-ant
velocity are averaged over the same time-interval $\Delta T(n)$.

\begin{figure}[tb]
\begin{center}
\includegraphics[width=0.35\textwidth]{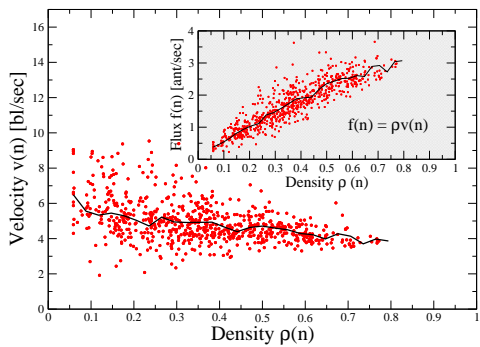}
\end{center}
\caption{Average velocity (solid line) and single-ant velocities (dots) for 
  unidirectional single-lane trail section of length $L=17~$bl. The
  corresponding flux-velocity relation, the so-called {\it fundamental
    diagram} is plotted in the inset. Mutual blocking is obviously
  suppressed as the average velocity is almost independent of the
  density. Consequently, the flux increases almost linearly with the
  density in the fundamental diagram (see inset).  }
\label{fig-fund}
\end{figure}

The average velocity of the ants is plotted against the corresponding
density in Fig.~\ref{fig-fund}; the resulting flow-density relation,
which is called {\em fundamental diagram} in traffic engineering
\cite{may,css}, is plotted in the inset of the Fig.~\ref{fig-fund}.
The most unusual feature of the data shown is that, unlike vehicular
traffic, there is no significant decrease of the average velocity with
increasing density \cite{may,kerner,css}. Consequently, the flux
obtained by the hydrodynamic equation increases approximately linearly
over the entire regime of observed density.  The jammed branch of the
fundamental diagram, which is commonly observed in vehicular traffic
and which is characterized by a monotonic decrease of flow with
increasing density, is completely missing 
in Fig.~\ref{fig-fund}.  Obviously effects of
mutual blocking, which are normally expected to become dominant at
high densities \cite{may,kerner,css}, are strongly suppressed in
ant-traffic.

\begin{figure}[tb]
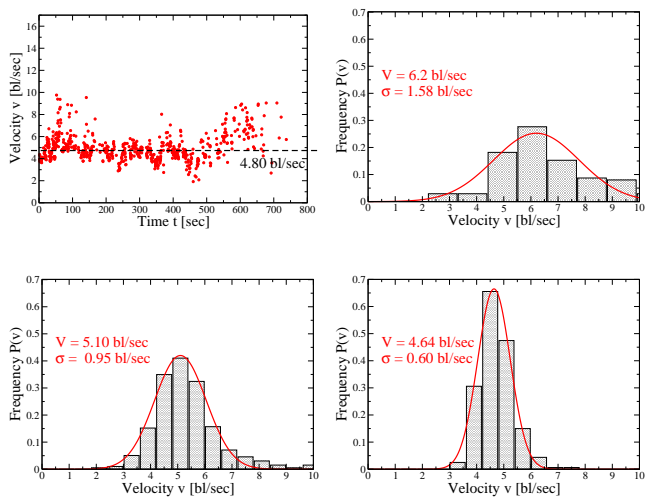

\begin{center}
\includegraphics[width=0.23\textwidth]{john-fig4a.eps}\hglue6pt
\includegraphics[width=0.23\textwidth]{john-fig4b.eps}\\ \vglue13pt
\vglue-8pt
\vspace{0.4cm}
\includegraphics[width=0.23\textwidth]{john-fig4c.eps}\hglue6pt
\includegraphics[width=0.23\textwidth]{john-fig4d.eps}
\end{center}
\caption{The velocities of individual ants are plotted as function 
  of time (top left). Furthermore, the velocity
  distributions of the ants for small ($\rho(n)\in[0,0.2]$, top right),
  intermediate ($\rho(n)\in[0.2,0,4]$, bottom left) and large
  ($\rho(n)\in[0.4,0.8]$, bottom right) densities are shown. For all regimes
  the corresponding Gaussian fit $P(v) =
  \frac{1}{\sigma\sqrt{2\pi}}\exp\left( -(V-v)^2/(2\sigma^2) \right) $
  is also shown (solid line).  }
\label{fig-uni-vel-dst}
\end{figure}

From the time-series of the single-ant velocities we have also determined 
their distributions in different density regimes (Fig.~\ref{fig-uni-vel-dst}).
The most striking feature is that the distribution becomes much
sharper with increasing global density whereas the most probable 
velocity decreases only slighty.  

\begin{figure}[tb]
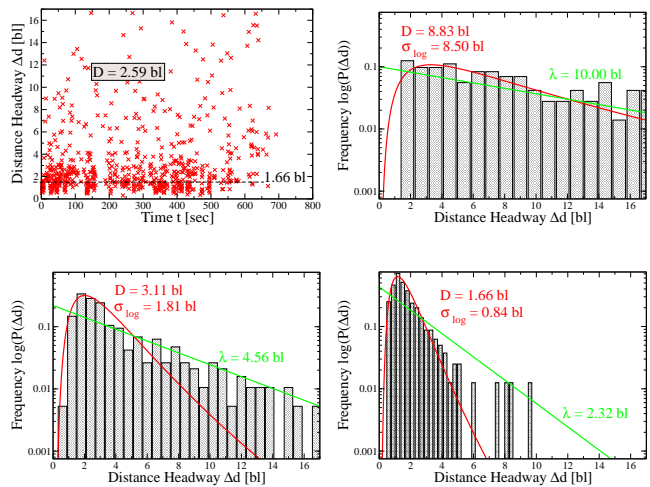

\begin{center}
\includegraphics[width=0.23\textwidth]{john-fig5a.eps}\hglue6pt
\includegraphics[width=0.23\textwidth]{john-fig5b.eps}\\ \vglue13pt
\vglue-8pt
\vspace{0.4cm}
\includegraphics[width=0.23\textwidth]{john-fig5c.eps}\hglue6pt
\includegraphics[width=0.23\textwidth]{john-fig5d.eps}
\end{center}
\caption{Distance-headways vs.\ time shows (top left)
  clustering of short distances which indicates the existence of a
  density independent intra-platoon separation between the ants. 
  The remaining three figures show the distance-headway distributions for
  small ($\rho(n)\in[0,0.2]$, top right), intermediate
  ($\rho(n)\in[0.2,0.4]$, bottom left) and large
  ($\rho(n)\in[0.4,0.8]$, bottom right) densities. Depending on the
  density regime a log-normal $P(\Delta d) =
  \frac{1}{\sqrt{2\pi}\sigma_\text{log}\Delta d }\exp\left(
    -(D-\log(\Delta d))^2/ (2\sigma_\text{log}^2) \right) $ or
  negative-exponential $P(\Delta d) = \exp\left(-\Delta d/\lambda
  \right) $ distribution applies.  }
\label{fig-uni-ddst}
\end{figure}

Another important quantity that characterizes the spatial distribution 
of the ants on the trail is the distance-headway distribution. 
The time-series (Fig.~\ref{fig-uni-ddst} top left), obtained by using 
(\ref{dh-def}), shows clustering of small distance-headways whereas 
larger headways are much more scattered. The distribution of these 
headways (see Fig.~\ref{fig-uni-ddst}) becomes much sharper with 
increasing density while the maximum shifts only slightly to smaller 
headways. At low densities, predominantly large distance-headways are 
found; the corresponding distribution for sufficiently long 
distance-headways is well described by a negative-exponential 
distribution which is characteristic of the so-called random-headway 
state \cite{may}. In contrast, at very high densities mostly very short 
distance-headways are found; in this regime, the log-normal distribution 
appeares to provide the best fit to our empirical data. 

The absence of a jammed phase in the fundamental diagram is closely 
related to the characteristic features of the distributions of the 
distance-headways of the ants along the observed section of the trail. 
The dominant, and directly observable, feature of this spatial 
distribution is the platoons formed by the ants. Ants inside a platoon 
move with almost identical velocities maintaining small distance-headways. 
These intra-platoon distance-headways are responsible for the clustering 
of data observed in the corresponding time-series Fig.~\ref{fig-uni-ddst}. 
In contrast, larger distance-headways are inter-platoon distances. The 
full distribution of distance-headways has an average of $D=2.59~$bl 
which is quite close to the value $D=1.66~$bl found for very high 
densities. This indicates the existence of a density-independent 
distance-headway for the ants moving inside platoons. Formation of these 
platoons has been demonstrated by our earlier simple models of ant-traffic 
\cite{chow,nishi,kunwarjap,kunwarjstat,johnjtb} which are appropriate 
extensions of the totally asymmetric simple exclusion process \cite{schuetz}.

The interpretations of the observed trends of variations of the flux, 
average velocity and distance-headway distribution with increasing 
density is consistent with the corresponding variation of the 
distribution of the velocities of the ants (Fig.~\ref{fig-uni-vel-dst}).
Ants within a platoon move at a slower average velocity whereas solitary 
ants can move faster if they detect a strong pheromone trace created by 
a preceeding platoon. Moreover, since fluctuations of velocities of 
different platoons are larger than the intra-platoon fluctuations, the 
distribution becomes sharper at higher densities because the platoons 
merge thereby reducing their number and increasing the length of the 
longest one. As can be seen in Fig.~\ref{fig-uni-vel-dst} the maximum of 
the velocity distribution is almost independent of the density. Its 
position at sufficiently large densities can be interpreted as platoon 
velocity, $v_p \approx 4.6~$bl/s.

In this letter, we have have reported results of our empirical studies 
of {\em unidirectional} ant-traffic on a natural ant trail. Remarkably, we 
have not observed any event of overtaking of one ant by another. Instead, 
we found formation of platoons which has been predicted by
simple models of ant-traffic under various conditions 
\cite{chow,nishi,kunwarjap,kunwarjstat,johnjtb}.  
In contrast to what has been previously been observed in 
various transport systems, especially vehicular traffic,
in ant-traffic (at least for the ant species 
and the trail systems used in our studies), flow always
increases monotonically with the density of the ants. In other words, 
no jammed branch is exhibited by the flow-density relation 
for ant-traffic. 

In highway traffic, the average velocity of the vehicles is
constant only for low densities, but 
the physical origin of this regime is
very different from the constant velocity of ants in ant-traffic. 
In low-density limit of vehicular traffic
the vehicles are well separated from each other and, therefore, can move 
practically unhindered in the so-called free-flow state.  On the other hand, 
in ant-traffic, this constant velocity regime is a reflection of the fact
that ants march together collectively forming platoons which reduce
the effective density.  We have also not found any evidence for
phenomena like hysteresis, synchronised flow, etc. which have been
reported for vehicular traffic \cite{css,kerner}. Although platoon
formation is considered to be relevant also for synchronized flow
\cite{kerner}, there is no characteristic velocity of these platoons,
in contrast to the case of ant traffic.

Thus, in spite of some superficial similarities, the characteristic
features of ant-traffic seems to be very different from those of
vehicular traffic and other typical transport systems. 
Perhaps, ant-traffic is analogous to human
pedestrian traffic \cite{BursteddeKSZ01,duss05,john-swarm}, as was
conjectured beautifully by H\"olldobler and Wilson in their classic
book \cite{holl90}; in a future work, we intend to explore this
analogy empirically and by quantitative modeling. Our results may have
important implications for swarm intelligence \cite{intelligence} and
ant-based computer algorithms \cite{bonabeau}.

We thank R. Gadagkar, T. Varghese, M. Kolatkar, P. Chakroborty and M. Burd 
for useful discussions, and D. Chaudhuri for a critical reading of the 
manuscript. AJ's field work in India has been supported by the 
German Academic Exchange Service (DAAD). DC acknowledges support from CSIR  
(India). 


\end{document}